\documentstyle[aps,floats,graphicx]{revtex}

\begin{document}

\newcommand{\bec}{\begin{center}}
\newcommand{\ec}{\end{center}}
\newcommand{\be}{\begin{equation}}
\newcommand{\ee}{\end{equation}}
\newcommand{\beqn}{\begin{eqnarray}}
\newcommand{\eeqn}{\end{eqnarray}}
\newcommand{\bet}{\begin{table}}
\newcommand{\ent}{\end{table}}
\newcommand{\bib}{\bibitem}

\wideabs{

\title{
Fingerprint of super-interdiffusion: anomalous intermixing in Pt/Ti
}

\author{P. S\"ule, M. Menyh\'ard} 
  \address{Research Institute for Technical Physics and Material Science,
www.mfa.kfki.hu/$\sim$sule, sule@mfa.kfki.hu\\
Konkoly Thege u. 29-33, Budapest, Hungary\\
}

\date{\today}

\maketitle

\begin{abstract}
The ion-sputtering induced transient enhanced intermixing has been studied by molecular dynamics (MD) simulations
in Pt/Ti and its anomalous nature has been explained as a superdiffusive transient enhanced interdiffusion.
We find ballistic mixing and a robust mass effect in Pt/Ti. The sum of the square of atomic displacements
($\langle R^2 \rangle$)
asymptotically grows nonlinearily and scales as $N^2$ and $\sim t^2$, where $N$ and $t$ are the ion-number fluence and the time of ion-sputtering, respectively.
This anomalous behavior explains the high diffusity tail in the concentration profile obtained
by 
Auger electron spectroscopy depth profiling (AES-DP) analysis in Pt/Ti/Si substrate (Pt/Ti) multilayer.
In  Ta/Ti/Pt/Si multilayer we find a linear time scaling of $\langle R^2 \rangle \propto t$ at the Ti/Pt interface indicating
the suppression of superdiffusive features.
We propose a qualitative explanation based on the accelerative effect of nonlinear forces provbided by the
anharmonic host lattice.

\end{abstract}
}

\section{Introduction}

  The most well known atomic diffusion mechanisms (simple site exchange, vacancy, self-interstitial mediated or concentration dependent) are thermally activated
processes \cite{Michely}.
However, there are a growing number of evidences are emerged 
that the anomalous broadening of interfaces or the high diffusity tail in the impurity concentration profile
could be the fingerprint of anomalously fast impurity diffusion
\cite{Abrasonis,Buchanan,Parascandola}.
It has been concluded by Abrasonis {\em et al.} that anomalously long diffusion depths
can only be understood by the assistance of lattice accelerative effects \cite{Abrasonis}.
Transient enhanced diffusion has also been reported in post-annealed
dopant implanted semiconductors \cite{Delugas,Melis,Solmi,Sparks,Venezia,Slijkerman}.

 Anomalously long interdiffusion dephts has been found in various diffusion
couples \cite{Abrasonis,Buchanan,Delugas,Melis,Solmi,Sparks,Venezia,Slijkerman}.
It has been reported that
anomalous interdiffusion is not driven by bulk diffusion parameters nor by thermodynamic forces (such as heats of alloying) \cite{Buchanan} or nor by heats of mixing \cite{Sule_NIMB04,Sule_PRB05}.
Transient intermixing (IM, surface alloying) has also been found the most recently by molecular dynamics simulations at cryogenic temperatures
during the deposition of Pt on Al(111)
\cite{Sule_condmat}.

 Computer simulations have revealed that
the mass-anisotropy of metallic bilayers governs ion-bombardment induced enhancement of interdiffusion at the interface \cite{Sule_PRB05} and
greatly influences surface morphology development 
\cite{Sule_SUCI,Sule_NIMB04_2} and intermixing properties during ion-sputtering \cite{Sule_NIMB04,Sule_PRB05}.
In mass-anisotropic metallic bilayers, such e.g. Al/Pt, or in Ti/Pt low-energy ion-bombardment leads also to
anomalously large broadening at the interface \cite{Sule_NIMB04,Sule_PRB05,Sule_SUCI}.
The intermixing length (and the mixing efficiency) scales nonlinearily with the mass anisotropy (mass ratio)
\cite{Sule_PRB05}. 
These enhanced atomistic transport processes cannot be understood by the established mechanisms of radiation-enhanced
diffusion (RED) \cite{Abrasonis}.
This is because RED cannot cause anomalously large diffusion depths (far beyond the ion penetration depth) \cite{Abrasonis}.  
Moreover, the observed asymmetry of intermixing in Ti/Pt and in Pt/Ti can not be understood
by the thermal spike model nor by RED \cite{Sule_condmat06b}.
Also, one might not explain the observed large intermixing length \cite{Buchanan} and
abrupt transport processes \cite{Sule_condmat} by
thermally activated diffusion. 

 It has already been shown recently that the asymmetry of intermixing has been found
in Pt/Ti and in Ti/Pt bilayers. However, no explanation is found for the very strong
and much weaker interdiffusion in Pt/Ti and in Ti/Pt, respectively \cite{Sule_condmat06b}.
   We propose in this article to understand the enhanced interdiffusion in Pt/Ti  as a superdiffusive atomic transport
process
since they fulfill the most important condition of superdiffusion:
the square of atomic displacements scales nonlinearily with the time of diffusion \cite{anomalous}.
We present evidences in this paper for the fullfillment of this condition for Pt/Ti.
 We would like to show that ballistic (athermal) intermixing 
occurs in Pt/Ti in contrast to Ti/Pt in which normal
radiation-enhanced thermally activated interdiffusion takes place.

\section{The experimental setup}

  Although the experimental setup has been outlined in ref. \cite{Sule_condmat06b},
however, we briefly also give the most important features of the
measuremnts.
  The experimental setup is the following:
According to the crossectional TEM (XTEM) results the thickness of the layers in
samples are:
Pt $13$ nm/Ti $11$ 
nm/Si substrate
(denoted throughout the paper as Pt/Ti), and Ta 21 nm (cap layer to prevent oxidation
of Ti)/Ti 11 nm/Pt 12 nm/Si substrate (denoted as Ti/Pt). 
The XTEM images are shown in Fig 1.
For the sake of simplicity we consider our multilayer samples as bilayers and we study the 
atomic transport processes at the Ti/Pt and Pt/Ti interfaces.
Both samples have been AES depth profiled by applying various sputtering conditions.
The sample has been rotated during sputtering. In the following we will outline results of $500$ eV
 Ar$^+$ ion bombardment at an
angle of incidence of $10^{\circ}$ (with respect to the surface).
The atomic concentrations of Pt, Ti, Ta and Si were calculated by the relative sensitivity method
taking the pure material's values from the spectra. The oxygen atomic concentration has been
calculated by normalizing the measured oxygen Auger peak-to-peak amplitude to TiO$_2$ 
\cite{Vergara}. The depth scale was determined by assuming that the sputtering yield ($Y_i$) in the
mixed layer is the weighted sum of the elemental sputtering yields ($\sum_i X_i Y_i$).
The broadening of the interface is frequently characterizied by the depth resolution. The depth resolution is defined  as the distance of points on the depth profile exhibiting 84 \% and  16 \% concentrations. This definiton has been introduced for the "normal" cases, when either the ion mixing or the roughening can be described by a Gaussian convolution resulting 
in an erf fuction transition in the depth profile. The same definition used for other cases as well, however.

  If the transition does differ from the $erf$ function (e.g. when the mobility of one of the components of a diffusion couple is much higher than that of the other's) one might give also the distances between points of $84$ \% and $50$ \%, and $50$ \% and $16$ \%.
The ratio of these distances gives us the asymmerty of intermixing (shown in Table 1).

\section{The setup of the atomistic simulations}

 In order to get more insight to the mechanism of interdiffusion 
 classical molecular dynamics simulations have also been used to simulate the ion-solid interaction
(using the PARCAS code \cite{Nordlund_ref}).
Here we only shortly summarize the most important aspects.
A variable timestep
and the Berendsen temperature control is used to maintain the thermal equilibrium of the entire
system. \cite{Allen}. The bottom layers
are held fixed in order to avoid the rotation of the cell.
Periodic boundary conditions are imposed laterarily and a free surface is left for the ion-impacts.
The temperature of the atoms in the
outermost layers was softly scaled towards the desired temperature to provide temperature control and ensure
that the pressure waves emanating from cascades were damped at the borders.
The lateral sides of the cell are used as heat sink (heat bath) to maintain the thermal equilibrium of the entire
system \cite{Allen}.
The detailed description of other technical aspects of the MD simulations are given in \cite{Nordlund_ref,Allen} and details specific to the current system in recent
communications \cite{Sule_NIMB04,Sule_PRB05,Sule_SUCI,Sule_NIMB04_2}.

  We irradiate the bilayers Pt/Ti and Ti/Pt 
with 0.5 keV Ar$^+$ ions repeatedly with a time interval of 10-20 ps between each of
the ion-impacts at 300 K
which we find
sufficiently long time for the termination of interdiffusion, such
as sputtering induced intermixing (ion-beam mixing) \cite{Sule_NIMB04}.
 The initial velocity direction of the
impacting ion was $10$ degrees with respect to the surface of the crystal (grazing angle of incidence)
to avoid channeling directions and to simulate the conditions applied during ion-sputtering. 
We randomly varied the impact position and the azimuth angle $\phi$ (the direction of the ion-beam).
In order to approach the real sputtering limit a large number of ion irradiation are
employed using automatized simulations conducted subsequently together with analyzing
the history files (movie files) in each irradiation steps.
In this article we present results up to 200 ion irradiation which we find suitable for
comparing with low to medium fluence experiments. 200 ions are randomly distributed
over a $20 \times 20$ \hbox{\AA}$^2$ area which corresponds to $\sim 5 \times 10^{15}$
ion/cm$^2$ ion fluence
and the removal of few MLs.

 The size of the simulation cell is $110 \times 110 \times 90$ $\hbox{\AA}^3$ including
57000 atoms (with 9 monolayers (ML) film/substrate).
At the interface (111) of the fcc crystal is parallel to (0001) of the hcp
crystal
and the close packed directions are parallel.
The interfacial system is a heterophase bicrystal and a composite object of
two different crystals with different
symmetry is created as follows:
the hcp Ti is put by hand on the (111) Pt bulk (and vice versa) and various structures are probed
and are put together randomly. Finally the one which has the smallest
misfit strain prior to the relaxation run is selected.
The difference between the width of the overlayer and the bulk does not exceed $2
-3$ \hbox{\AA}.
The remaining misfit is properly minimized below $\sim 6 \%$ during the relaxation
process so that the Ti and Pt layers keep their original crystal structure and we
 get an
atomically sharp interface.
During the relaxation (equilibration) process the temperature is softly scaled down
to zero.
According to our practice we find that during the temperature scaling down the 
structure
becomes sufficiently relaxed therefore no further check of the structure has been
 done.
Then the careful heating up 
of the system to $300$ K has been carried out. The systems were free from any serious built-in strain
and the lattice mismatch is minimized to the lowest possible level.
The film and the substrate are $\sim 20$ and $\sim 68$ $\hbox{\AA}$ thick, respectively.

 In order to reach the most efficient ion energy deposition at the interface,
we also initialize recoils placing the ion above the interface by $10$ \hbox{\AA} (and
below the free surface in the 9 ML thick film) also at grazing angle of incidence
($10^{\circ}$ to the surface)
with $500$ eV ion energy.
In this way  
we can concentrate directly on the intermixing phenomenon avoiding
many other processes occur at the surface (surface roughening, sputter erosion, ion-induced surface diffusion, cluster ejection, etc.) which weaken energy deposition at the interface.
Further simplification is that channeling recoils are left to leave the cell
and in the next step these energetic and sputtered particles are deleted.

 We used a tight-binding many body potential, developed by Cleri and
Rosato (CR) on the basis of the second moment approximation to the density of states \cite{CR}, to describe interatomic interactions.
This tight-binding approach is formally analogous to the embedded-atom method (EAM) \cite{EAM}.
This type of a potential gives a good description of lattice vacancies, including atomic migration
properties and a reasonable description of solid surfaces and melting \cite{CR}.
Since the present work is mostly associated with the elastic properties,
melting behaviors, interface and migration energies, we believe the model used should be suitable for this study.
The interatomic interactions are calculated up to the 2nd nearest neighbors
and a cutoff is imposed out of this
region.
This amounts to the maximum interatomic distance of $\sim 6$ \hbox{\AA}.
 For the crosspotential of Ti and Pt we employ an interpolation scheme \cite{Sule_PRB05,Sule_SUCI,ZBL}
between the respective elements.
The CR elemental potentials and the interpolation scheme for heteronuclear interactions
have widely been used for MD simulations \cite{Sule_PRB05,Stepanyuk2,Goyhenex,Levanov}.
 The Ti-Pt interatomic crosspotential of the Cleri-Rosato \cite{CR} type is fitted to the experimental
 heat of
mixing of the corresponding alloy system \cite{Sule_NIMB04,Sule_NIMB04_2}.
The scaling factor $r_0$ (the heteronuclear first neighbor distance) is calculated as the average of the elemental first neighbor distances.

 The computer animations can be seen in our web page \cite{web}.
Further details  are given in \cite{Nordlund_ref} and details specific to the current system in recent
communications \cite{Sule_PRB05,Sule_NIMB04}.

\section{Results}

  In Fig 3 the evolution of the sum of the square of atomic displacements (SD) of all intermixing atoms $\langle R^2 \rangle= \sum_i^N [{\bf r_i}(t)-{\bf r_i}(t=0)]^2$, obtained by molecular dynamics simulations, where (${\bf r_i}(t)$ is the position vector of atom 'i' at time $t$, $N$ is the total number of atoms included in the sum), can be followed as a function of the ion fluence. 
Lateral components ($x,y$) are excluded from 
$\langle R^2 \rangle$ and only contributions from intermixing atomic displacements perpendicular to the layers
are included ($z$ components). We follow during simulations the time evolution
of $\langle R^2 \rangle$ which reflects the atomic migration through the interface (no other
atomic transport processes are included).
Note, that we do not calulate the mean square of atomic displacements (MSD) which is an
averaged SD over the number of atoms included in the sum (MSD=$\langle R^2 \rangle/N$).
MSD does not reflect the real physics when localized events take place, e.g. when only few dozens of atoms
are displaced and intermixed.
In such cases the divison by $N$, when $N$ is the total number of the atoms in the simulation cell
leads to the meaningless $\langle R^2 \rangle/N \rightarrow 0$ result when $N \rightarrow \infty$,
e.g. with the increasing number of atoms in the simulation cell.
Also, it is hard to give the number of "active" particles which
really take place in the transient atomic
transport processes.
\begin{figure}[hbtp]
\begin{center}
\includegraphics*[height=6cm,width=6.9cm,angle=0.]{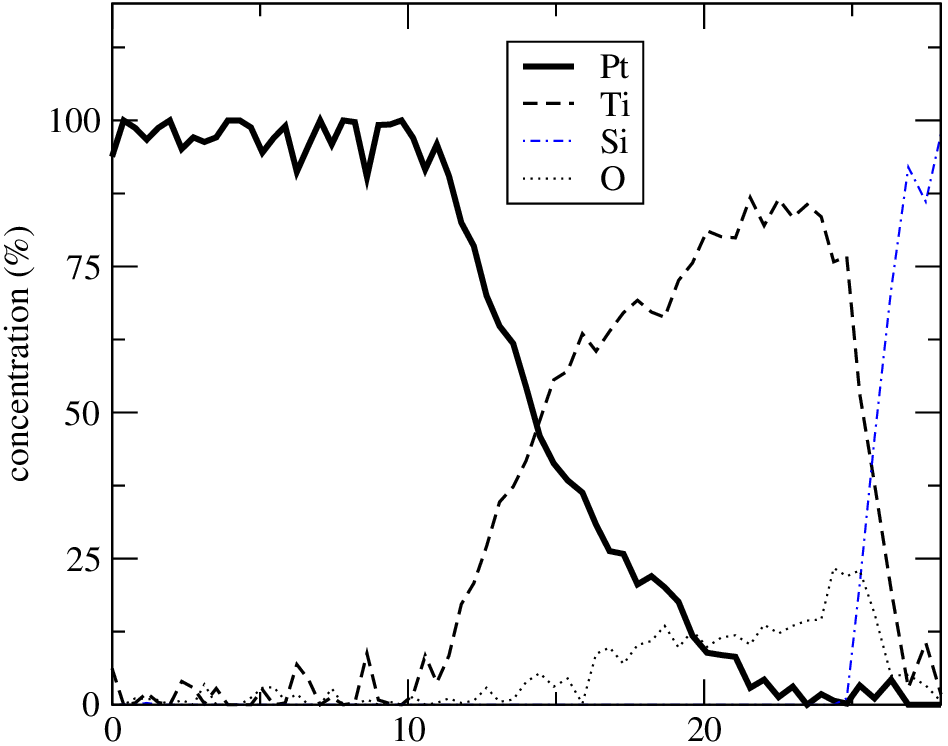}
\includegraphics*[height=6cm,width=6.9cm,angle=0.]{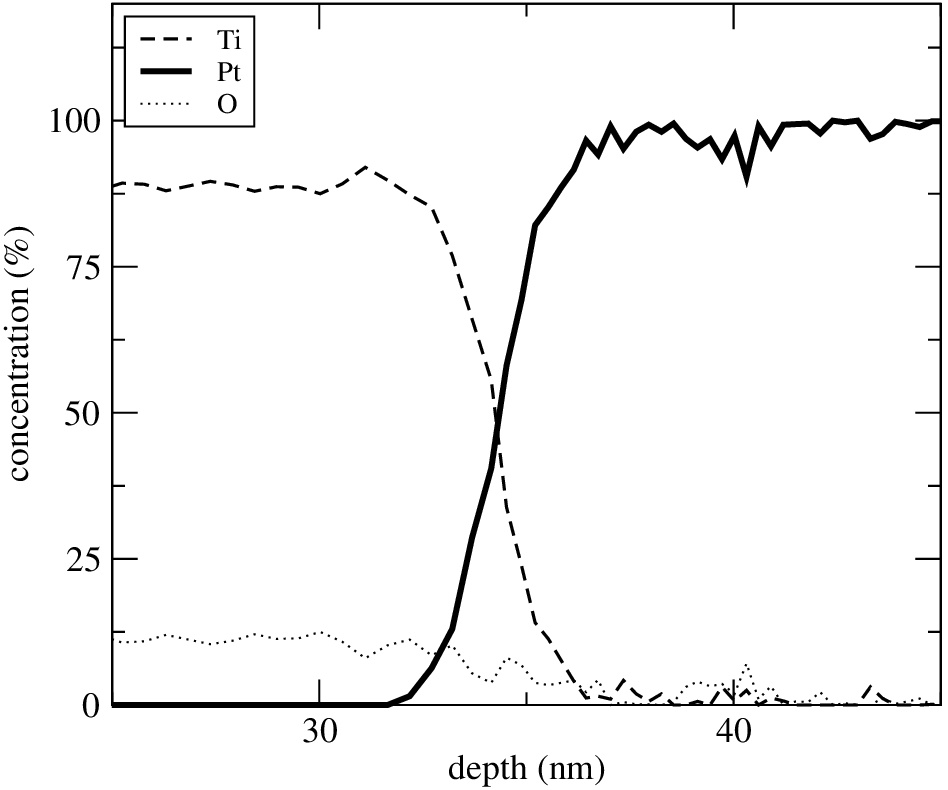}
\caption[]{
The concentration depth profile as a function of the removed layer thickness (nm)
 obtained by AES depth profiling analysis
using ion-sputtering at 500 eV Ar$^+$ ion energy ($10^{\circ}$ grazing angle of incidence with respect to the surface) in P
t/Ti (upper Fig: 1a) and Ti/Pt (lower Fig: 1b).
}
\label{fig2}
\end{center}
\end{figure}
Hence we prefer to use the more appropriate quantity SD.
In Fig 3 we present
$\langle R^2 \rangle$ as a function of the number of ion impacts $N_i$ (ion number fluence).
$\langle R^2 \rangle (N_i)$ corresponds to the final value of
$\langle R^2 \rangle$
obtained during the $N_i$th simulation. The final relaxed structure of the simulation of the
$(N_i-1)$th ion-bombardment is used as the input structure for the $N_i$th ion-irradiation.
The
asymmetry of
 mixing can clearly be seen when $\langle R^2 \rangle (N_i)$
and the depth profiles are compared in 
Ti/Pt and in Pt/Ti in Figs 1a,1b and 2.
The computer animations of the simulations together with the plotted broadening values
at the interface in inset
Fig 3 of ref. \cite{Sule_condmat06b} also reveal the stronger
interdiffusion in Pt/Ti \cite{web}.

 As it has already been shown in ref. \cite{Sule_condmat06b}
 a relatively weak intermixing is found in Ti/Pt ($\sigma \approx 20$ $\hbox{\AA}$) while an unusually high
interdiffusion occurs in the Pt/Ti bilayer ($\sigma \approx 70$ $\hbox{\AA}$).
MD simulations provide $\sim 8$ ML ($\sim 20$ $\hbox{\AA})$ and $\sim 16$ ML ($\sim 40$ $\hbox{\AA})$ thick interface after $200$ ion impacts, respectively.
The comparison of the measured $\sigma$ with the simulated broadening
using the $84-16$ \% rule in both cases can only be carried out with great care
\cite{Sule_condmat06b}.

  Moreover, the applied setup of the simulation cell, in particular the $20$ $\hbox{\AA}$ film thickness is assumed to be
appropriate for simulating broadening.
Our experience shows that the variation of the film thickness does not affect the
final result significantly, except if ultrathin film is used (e.g. if less than
$\sim 10$ $\hbox{\AA}$ thick film). At around $5$ or less ML thick film surface roughening could affect
mixing, and vice versa \cite{Sule_NIMB04_2}.
Also, we do not carry out complete layer-by-layer removal as in the experiment.
It turned out during the simulations that the ions mix the interface the most efficien
tly
when initialized $\sim 10 \pm 3$ $\hbox{\AA}$ above the interface.
This value is naturally in the range of the projected range of the ions.
Hence, the most of the broadening is coming from this regime of ion-interface distance
.
Initializing ions from the surface it takes longer and more ions are needed to
obtain the same level of damage and broadening at the interface as it has been found b
y ion-bombarding
from inside the film.

 In ref. \cite{Sule_condmat06b} we have shown that 
  the measured ion-sputtering induced broadening of $\sigma 
\approx 20$ $\hbox{\AA}$ for Ti/Pt is in nice agreement with
the simulated value.
Hence we expect that the most of the measured $\sigma$ is coming from intermixing
and interface roughening contributes to $\sigma$ only slightly.
The nice agreement could be due to that the saturation of intermixing (broadening) dur
ing ion-sputtering is insensitive to
the rate of mixing in the as received samples
($\sigma_0 \approx 15$ $\hbox{\AA}$ interface width (including the interface roughenin
g) in both samples).
This is because
a sharp interface and a weakly mixed one (before bombardment in the as-received samples) lead to the same magnitude of broadening
upon ion-sputtering, because the binary systems reach the same steady state
of saturation under the same conditions (ion energy, impact angele, etc.).
This is rationalized by our finding that
during simulations we start from a sharp interface ($\sigma_0 \approx 0$) and
we get a very similar magnitude of $\sigma$ than by AES.



\section{Discussion}

\subsection{The nonlinear scaling of $\langle R^2 \rangle$}

  In recent papers we explained single-ion impact induced intermixing as an interdiffusion
process governed by the mass-anisotropy parameter (mass ratio) in these bilayers
\cite{Sule_PRB05,Sule_SUCI}.
We also present $\langle R^2 \rangle$ results for Cu/Co bilayer in Fig 2.
Ti/Pt and Pt/Ti are mass-anisotropic, while Cu/Co is
a mass-isotropic bilayer.
Cu/Co can be used as a reference system, since it is composed of immiscible
elements, hence a weak rate of intermixing is expected for.
  It has already been shown in refs. \cite{Sule_NIMB04,Sule_PRB05} that the backscattering of the hyperthermal particles (BHP)
at the heavy interface leads to the increase in the energy density of the collisional cascade.
\begin{figure}[hbtp]
\begin{center}
\includegraphics*[height=6cm,width=7.5cm,angle=0.]{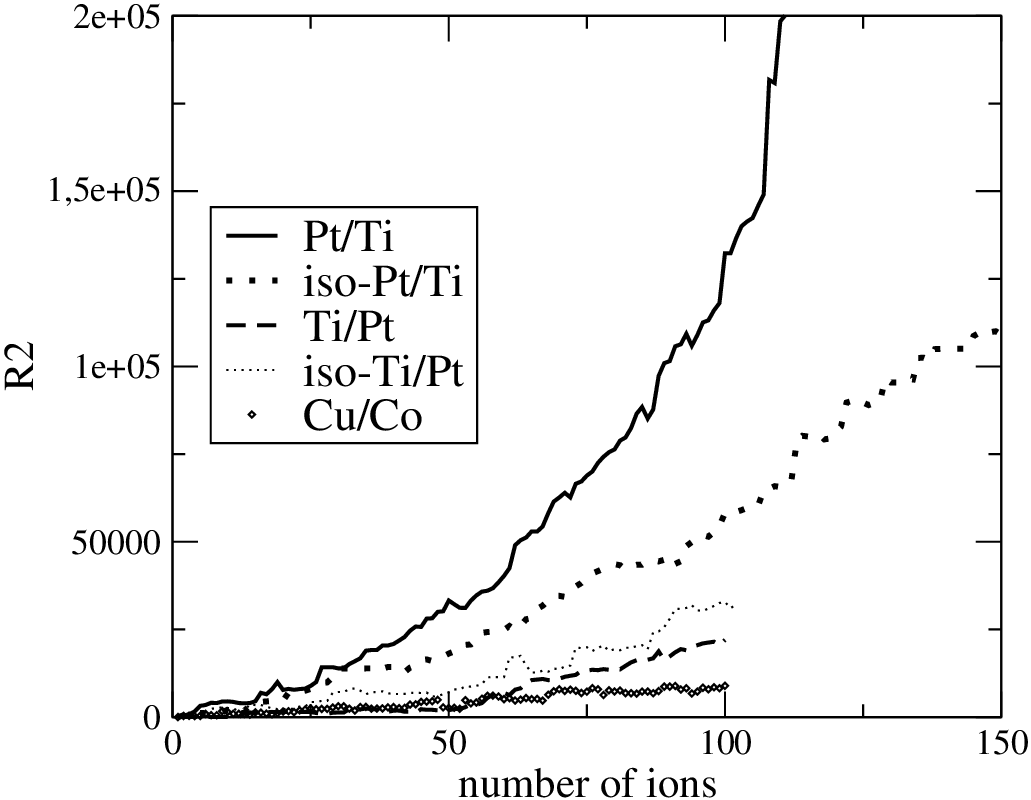}
\caption[]{
The simulated square of intermixing atomic displacements $\langle R^2 \rangle$ ($\hbox{\AA}^2$) in Pt/Ti, Ti/Pt and in Cu/Co as a function of the
ion-fluence (number of ions) obtained during the ion-sputtering of these bilayers at 500 eV ion energy (results are shown up to 100 ions).
The dotted lines (iso-Pt/Ti and iso-Ti/Pt) denote the results obtained for the artificial mass-isotropic
Pt/Ti and Ti/Pt bilayers, respectively.
The ions are initiated from the surface.
{\em Inset}:
The simulated broadening at the interface in $\hbox{\AA}$ as a function of
the number of ions at 500 eV ion energy.
The ions are initiated from $10$  $\hbox{\AA}$ above the interface.
The error bars denote a statistical uncertainty in the measure of broadening.
}
\label{fig3}
\end{center}
\end{figure}
We have found that the jumping rate of atoms through the interface is seriously affected by the
mass-anisotropy of the interface when energetic atoms (hyperthermal particles) are present and which leads to the
preferential intermixing of Pt to Ti \cite{Sule_NIMB04,Sule_PRB05}.

  According to the BHP model,
  in Pt/Ti, the energetic Ti light particles are backscattered downwards and confined
below the heavy interface
in the bulk which results in increasing energy density below the interface.
In Ti/Pt we find the reversed case: the Ti atoms are backscattered upwards, towards the
nearby free surface (the ion-sputtered surface is always close to the interface when the
interface is irradiated).
The injection of the Pt particles to the Ti phase can be seen in lower Fig. 4.
We know from ref. \cite{Sule_NIMB04} that in Ti/Pt thermal spike occurs also at low ion energies.
At $0.5$ keV energy we also find thermal spike with the average lifetime of $\tau \approx 5$ ps averaged
during few tens of consequtive events.
This is due to the backscattering of light particles at the interface as it has been shown
in refs. \cite{Sule_PRB05,Sule_NIMB04}.
Hence mass anisotropy enhances thermal spike in both bilayers.
In Pt/Ti, however we find the further enhancement of the heat spike and
$\tau \approx 10$ ps.
Unfortunatelly we cannot explain further enhancement of interdiffusion 
in Pt/Ti by simple mass effect.
We reach the conclusion that there must be another accelerative field which
speeds up Pt particles to ballistic transport.
This is reflected by the divergence of $\langle R^2 \rangle$ from linear scaling in Fig 3. for
Pt/Ti.
Interdiffusion takes place via ballistic jumps (ballistic mixing), when $\langle R^2 \rangle$ grows
asymptotically as $N^2$, where $N$ is the number fluence
(the same asymptotics holds as a function ion-dose or ion-fluence).
This can clearly be seen in Fig. 3 for Pt/Ti.
The horisontal axis is proportional to the time of ion-sputtering, hence
$\langle R^2 \rangle 
\propto t^2$ which is the time scaling of ballistic atomic transport
\cite{anomalous}.
In our particular case $1$ ion-bombardment corresponds
to $t \sim 10$ ps which we find sufficient time for the evolution of $\langle R^2 \rangle$.
Anyhow, above this $t$ value the asymptotics of $\langle R^2 \rangle (t)$ is invariant
to the choice of the elapsed time/ion-bombardment induced evolution of $\langle R^2 \rangle (t)$.
Hence the transformation between ion-fluence and time scale is allowed.
$\langle R^2 \rangle \propto t^2$ and $\langle R^2 \rangle \propto t$
time scalings have been found even for the single-ion impacts averaged for few events (when $\langle R^2 \rangle (t)$ is plotted only for a single-ion impact) for Pt/Ti and Ti/Pt, respectively.

  No such ballistic behavior can be seen for Ti/Pt in Fig 3.
In Ti/Pt we find $\langle R^2 \rangle \propto t$ time scaling which is
due to the shorter average lifetime of the collisional cascades \cite{Sule_PRB05,Sule_NIMB04}.
The mean free path of the energetic particles are much shorter in Ti/Pt (this can be seen
qualitatively in Fig 4. if we compare the length of the atomic trajectories for Pt between
upper and lower panels of Fig 4.).
The $\langle R^2 \rangle \propto t^2$ scaling used to be 
considered as the signature of
anomalous diffusion (superdiffusion) in the literature \cite{anomalous}.
In the upper panel of Fig 2. the fingerprint of superdiffusive feature
of intermixing is detected by AES as a tail in the concentration profile of Pt.
No such tail occurs for Ti/Pt in the lower panel of Fig 2. where the profile
can be characterized by  "normal" $erf$ functions.
Superdiffusive features, however, have never been reported before for intermixing, only
for e.g. random walk of adatoms (Levy flight) on solid surfaces \cite{anomalous}.
Therefore, the Pt/Ti system is highly unusual and there has to be more surprise in store.
The anomalous impurity diffusion of N in stainless steel \cite{Abrasonis} and 
the observed large interdiffusion depths in various transition metal/Al diffusion
couples \cite{Buchanan} could also be understood as super-interdiffusive processes.
\begin{figure}[hbtp]
\begin{center}
\includegraphics*[height=5.5cm,width=6.9cm,angle=0.]{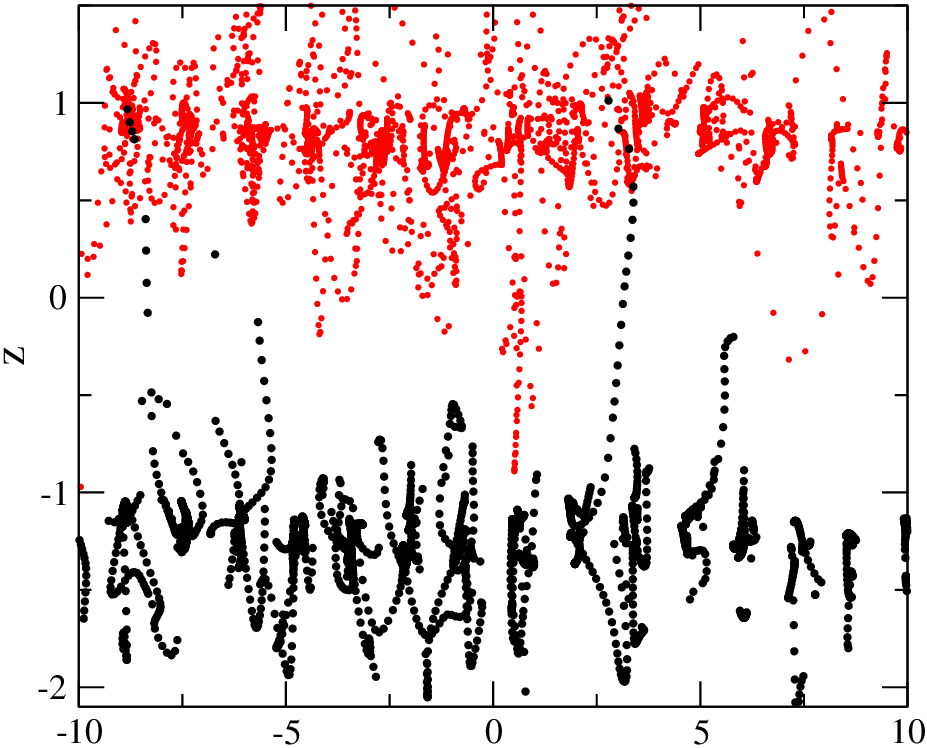}
\includegraphics*[height=5.5cm,width=6.9cm,angle=0.]{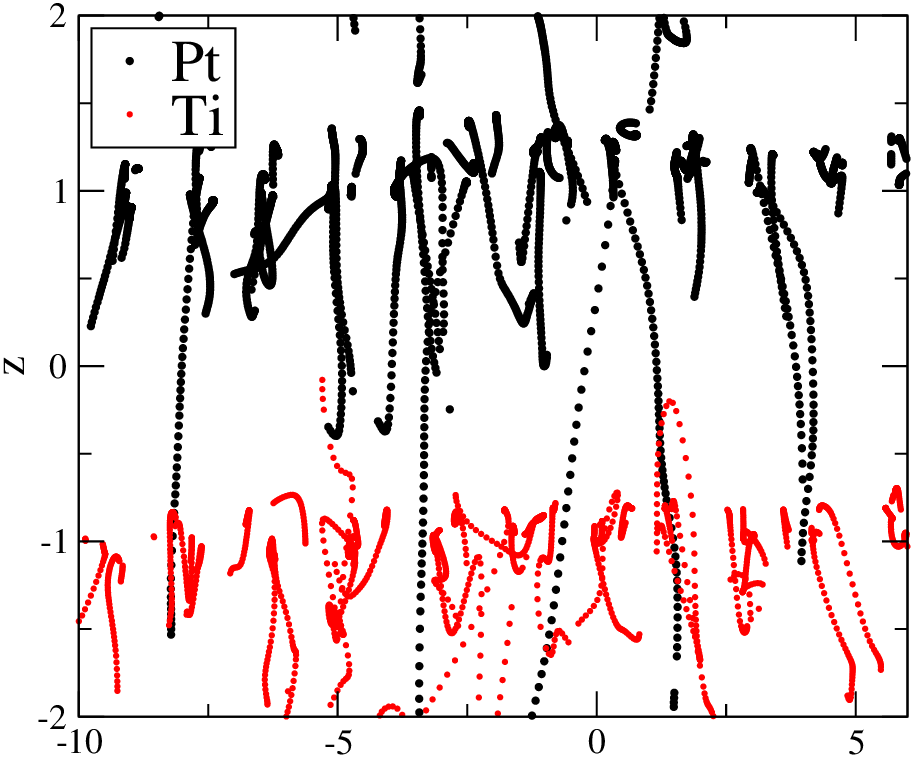}
\caption[]{
The crossectional view of a typical collisional cascade at the interface
with atomic trajectories (two monolayers are shown at the interfaces as a crossectional
slab cut in the middle of the simulation cell)
in Ti/Pt (upper panel) and in Pt/Ti (lower panel).
The positions of the energetic particles are collected up to 500 fs during
a $500$ eV single ion-impact event initialized $10$ $\hbox{\AA}$ above the interface
using the same conditions as in other simulations.
The vertical axis corresponds to the depth position given in $\hbox{\AA}$.
The position $z=0$ is the depth position of the interface.
}
\label{fig4}
\end{center}
\end{figure}
 The trajectories of the reversed Ti recoils can be seen in upper Fig. 4 and
no intermixing Ti atomic positions can be found in upper panel of Fig 4 (Ti/Pt).
Although, Fig. 4 has no any statistical meaning (as it should be), however,
the trajectories are selected among typical events hence some useful information
can be obtained for the transport properties of energetic Pt atoms.
In lower panel of Fig 4. we can see the ballistic trajectories of intermixing hyperthermal Pt atoms (Pt/Ti).
The reversed fluxes of the Ti particles at the interface
and the weaker intermixing of Pt atoms to the Ti phase
result in
the weaker intermixing in Ti/Pt than
in the Pt/Ti system.
This is nicely reflected in the $\langle R^2 \rangle$ of Ti/Pt which remains in the range of Cu/Co.
Hence Fig. 4 depicts us at atomistic level what we see in the more statistical quantity 
$\langle R^2 \rangle$. In Ti/Pt we see much shorter inter-layer atomic trajectories
while in Pt/Ti ballistic trajectories of Pt atoms can be seen (going through the interface).

\begin{figure}[hbtp]
\begin{center}
\includegraphics*[height=6cm,width=6.9cm,angle=0.]{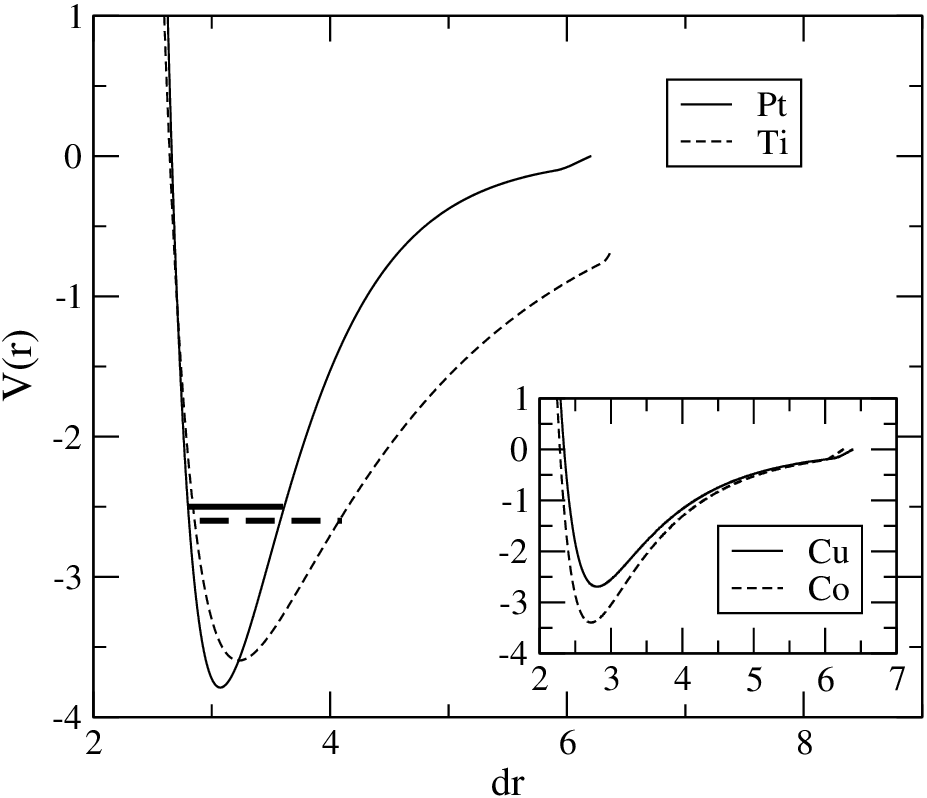}
\caption[]{
The Cleri-Rosato potential energy (eV) as a function of the
interatomic distance ($\hbox{\AA}$) in Ti and in Pt averaged for
an atom with its 12 first neighbors in the hcp and fcc lattices, respectively.
}
\label{fig5}
\end{center}
\end{figure}

\subsection{The effect of mass and size anisotropy}

 In order to clarify the mechanism of intermixing and to understand how much
the interfacial anisotropy influences IM
  simulations have been carried out with mass ratio $\delta$ is artificially set to $\delta
 \approx 1$ (mass-isotropic, ions are initiated from the surface). We find that $\langle
R^2 \rangle$ is below the corresponding curve of Pt/Ti (see
Fig 3, iso-Pt/Ti, dotted line, and animation \cite{web}).
The $\langle R^2 \rangle$ scales nearly linearily as a function of the number of ions
(and with $t$) for iso-Pt/Ti.
The asymptotics of $\langle R^2 \rangle$ is sensitive to the mass ratio.
Hence we reach the conclusion that the mass-effect is robust,
and the energetic particle scattering mechanism is weakened
when
mass-anisotropy
vanishes and the magnitude of intermixing is also weakened.

 The same type of MD simulations has been carried out in Cu/Co
as well, and the results are shown in Fig 3. The Cu/Co system is mass-isotropic,
and thus we expect that the magnitude of $\langle R^2 \rangle$ obtained for  iso-Ti/Pt and iso-Pt/Ti should be 
similar to that of Cu/Co.
It turns out, however, that the $\langle R^2 \rangle$
of iso-Ti/Pt and Ti/Pt are higher than that of Cu/Co. The $\langle R^2 \rangle$
of iso-Ti/Pt is somewhat even higher than that of Ti/Pt.
Hence the effect of mass-isotropy on $\langle R^2 \rangle$ is opposite in Ti/Pt
than in Pt/Ti.
This might be due to the reversed fluxes of Ti recoils in the mass-anisotropic Ti/Pt,
shown in upper panel of Fig 4.

  To understand the physical origin of the higher intermixing in iso-Pt/Ti
than in Cu/Co,
we should consider a more complex anisotropy of the interface
then previously
expected. 
We can rule out the effect of thermochemistry (e.g. heat of mixing $\Delta H$) 
because
it has also been shown recently, that $\Delta H$ has no
apparent effect on intermixing during single-ion impact in this bilayer \cite{Sule_NIMB04}.
 Also, thermochemistry does not explain the mixing asymmetry observed in Ti/Pt and
in Pt/Ti since $\Delta H$ must be the same in the two systems.
Moreover, simulated ion-sputtering with $\Delta H \gg 0$ (switching off the attractive term of the cross-potential
and using a repulsive (Born-Mayer) Ti-Pt interaction potential for Pt/Ti)
leads to intermixing instead of the expected sharp interface (see also animations \cite{web}).
We reach the conclusion that there must be a strong accelerative effect in the system which even
counterbalance the effect of heteronuclear repulsion.

  The employed interaction potential energy functions are plotted
in Fig ~(\ref{fig5}) as a function of the interatomic distance for Ti and for Pt. It can clearly be seen that
the thermal atomic sizes (the width of the potential valley at higher atomic kinetic energy) are different.
This difference could not be realized during the short time period of a collisional cascade
($\tau < 0.5$ ps).
However, the increased lifetime of the thermal spike in these bilayers allows few vibrational periods
and which is sufficient for the onset of the effect of atomic size anisotropy.
The atomic radius (the width of the vibration around the equilibrium lattice site) of the Ti atoms
is larger than that of in Pt. The atomic size anisotropy is even more robust at 1-2 eV energies above the energy minimum (hyperthermal atomic energies, see the horizontal lines in Fig 4). This explains the fact that the smaller and heavier Pt atoms are the first 
ballistic diffuser during ion-beam mixing \cite{Sule_NIMB04}. 
The atomic size difference and the mass-anisotropy together could explain at least partly the stronger simulated (and measured) intermixing in Pt/Ti and in Ti/Pt than 
in Cu/Co. In the latter mass-isotropic bilayer no serious atomic size difference (anisotropy) is found (see the inset Fig ~(\ref{fig5})).
The mixing in the iso-Pt/Ti could also be understood by the size anisotropy:
the larger energetic Ti atoms are reversed at the interface and confined in the bulk.
The smaller energetic Pt atoms, however, are injected to the Ti bulk during
the cascade period.

\subsection{Nonlinear accelerative field}

 Since it is evident from the analysis of the previous subsections that mass and size
anisotropy do not fully acount for the enhancement of IM in Pt/Ti, there must be
another source(s) of the divergence of $\langle R^2 \rangle$.
The nearly parabolic time scaling of $\langle R^2 \rangle$ suggests ballistic
IM induced by intrinsic acceleration of IM Pt particles.
The following simple picture seems to explain the nonlinear time scaling
of $\langle R^2 \rangle$:
Instead of the rapid solidification of the thermal spike intrinsic lattice
effects speed up inter-layer transport of Pt atoms until
saturation occurs in the Ti phase.

 Normal (thermally activated) atomic transport processes have already been carefully ruled out
in ref. \cite{Abrasonis} for the anomlaous diffusion of N in nitrided austenitic stainless steel.
There is no specific reason to expect that these reasonings do not hold for Pt/Ti.
Namely, we rule out that vacancy, multivacancy or self-interstitial mediated, stress induced or concentration
dependent, thermal acceleration  diffusion could explain the occurence of the asymmetry and the anomalous
nature of interdiffusion in Pt/Ti.

 Although, the anisotropy present in the bilayer systems have serious effect on intermixing,
the asymmetry of IM still remains unexplained.
This seems to be puzzling at first sight, however, we do believe that similar nonlinear effects
are responsible for the large diffusion depth of Pt as it has been proposed by
Abrasonis {\em et al.} in ref. \cite{Abrasonis}.
Similar nonlinear forces induced accelerative mechanism has been found the most recently for other diffusion couples
\cite{Sule_condmat}.
On the basis of these findings
anharmonic lattice vibrations (intrinsic localized modes) could promote atomic mobility.
It has already been shown that the localization of vibrational energy within an extended lattice
can influence various materials properties \cite{Swanson}.

 Since the enchanced atomic mobility is also present in weakly repulsive heteronuclear field
(with repulsive Ti-Pt internuclear potential)
one might expect that the accelerative field is rather strong and could indeed be due to
the intrinsic localization of anharmonic vibration modes at the interface region.
The fundamental question remains to be answered yet that
if nonlinear and long-range forces drive superdiffusion,
why these nonlocal anharmonic vibrational modes are active in Pt/Ti and why they are suppressed in Ti/Pt.
Using repulsive Ti-Pt potential in Ti/Pt we find the further
suppression of IM hence no accelerative long-range forces are active in this
material which could not surpass the slow down of the particles in the
repulsive interparticle field.

 The asymmetry of accelerative effects has also been found in Pt/Al couple during simulated
vapor deposition of Pt on Al(111). No transient intermixing is found for Al deposition on Pt(111) \cite{Sule_condmat}.
In this case we explained the apparent asymmetry by atomic size mismatch \cite{Sule_condmat}.
The ultrafast diffuser specie must be the smaller one. Moreover the host material atoms must be not only
larger but also anharmonic. This is indeed also the case for Ti.
Hence anharmonic vibrational modes could be excited in the
anharmonic host material which exhibits accelerative field. 
In the reverse case the larger particles could not excite the nearly harmonic host material with smaller particles.

 It should also be noted that the intrinsic accelerative effects of the anharmonic lattice modes
accelerate moving particles during the cascade and thermal spike period increasing the lifetime of
the heat spike. Few of these accelerated particles could reach hyperthermal kinetic energies (hot atoms).
According to our simulations in Pt/Ti
the mean free path of these particles is in the range of $\sim 10$ $\hbox{\AA}$. 
More careful analysis certainly will be necessary, however, to identify the details of this
specific mechanism which could also hold for couple of other diffusion couples.

\section{Conclusions}

 We could reproduce by means of  MD simulations the experimentally observed mixing asymmetry between Ti/Pt and Pt/Ti.  
We find a robust mass effect on interfacial mixing in Pt/Ti, although
the mass effect does not fully account for intermixing in Pt/Ti:
the atomic size and mass-anisotropy are together could be responsible for
the observed strong interfacial mixing in Pt/Ti and
the weaker mixing in Cu/Co is explained by atomic size and mass-isotropy.
We get a nice agreement for interface broadening in Ti/Pt with experiment while for
Pt/Ti the discrepancy is not negligible.
We conclude from this that interface roughening might has a significant
contribution to broadening in Pt/Ti.

 We also find that the sum of the squares of atomic displacements ($\langle R^2 \rangle$) through the anisotropic interface
scales nonlinearily in Pt/Ti ($\langle R^2 \rangle \propto t^2$) as a function of the time (and the ion-number fluence)
as shown in Fig. 3.
 From the upper panel of Fig. 2 we see the experimental fingerprint of ballistic mixing (super-interdiffusion)
as a well developed tail in the concentration profile of Pt with an approximate length of $\sim 25$
$\hbox{\AA}$ along the depth direction.
The long range (high diffusity) tail in the AES spectrum provides a compelling evidence for the superdiffusive transport
process of Pt atoms.
In Ti/Pt a nearly linear scaling ($\langle R^2 \rangle \propto t$) is found which is due to the shorter average
lifetime of the collisional cascade period and which leads to
shorter mean free path of the energetic Pt particles.
The lack of
a tail in the AES concentration profile for Ti/Pt (lower panel of Fig 2.) indicates a
concentional (thermally activated) mechanism of interdiffusion in Ti/Pt.
In this system ion-sputtering increases broadening only slightly: in the as received sample
we find $\sigma_0 \approx 15$ 
$\hbox{\AA}$
, and $\sigma \approx 20$ $\hbox{\AA}$ is measured after AES depth profiling.
Based on atomistic simulations we find that
the backscattering of the sizeable and light Ti particles at the interface suppresses further
intermixing.

 We conclude that the observed and simulated long range depth distribution of Pt atoms
in the Ti phase of Pt/Ti cannot be understood by any established mechanisms of radiation-enhanced
diffusion.
 In Pt/Ti, we find that further accelerative effects could enhance interdiffusion.
A specific mechanism might come into play which speeds up Pt particles
in the Ti bulk or at the interface.
The divergence of $\langle R^2 \rangle$ clearly indicates that accelerative effects are present
in the lattice.
At low ion energies, such as 0.5 keV, the damage energy
caused by the slowed down ions cannot accelerate particles beyond the
ion penetration depth.
Moreover, the observed asymmetry also could not be understood within a simple
picture of collisional cascades. 
The mean diffusional depth could reach few tens of $\hbox{\AA}$, which
is beyond the mean free path of recoils at such a low ion energies.
We support the explanation raised in ref. \cite{Abrasonis} that nonlinear
forces could govern the accelereration of the particles in the bulk.
However, it is unclear which possible scenario (breather-like solitons, anharmonic longitudinal vibrations, etc.) explain the superdiffusive mechanism \cite{Abrasonis}.

{\scriptsize
This work is supported by the OTKA grant F037710
from the Hungarian Academy of Sciences.
We wish to thank to K. Nordlund 
for helpful discussions and constant help.
The work has been performed partly under the project
HPC-EUROPA (RII3-CT-2003-506079) with the support of
the European Community using the supercomputing 
facility at CINECA in Bologna.
The help of the NKFP project of 
3A/071/2004 is also acknowledged.
}

\end{document}